\documentclass{aa}
\usepackage{latexsym}
\usepackage{amsmath}
\usepackage{amssymb}
\usepackage{graphicx}
\usepackage[english]{babel}
\title{Stellar yields with rotation and their effect on
chemical evolution models}
\author{C. Chiappini\inst{1}, F. Matteucci\inst{2,1}  \and G. Meynet \inst{3}}
\institute{Osservatorio Astronomico di Trieste, Via G. B. Tiepolo 11,\newline I - 34131 Trieste, Italia \and Dipartimento di Astronomia, Universita' degli Studi di Trieste,\newline Via G. B. Tiepolo 11, I - 34131 Trieste, Italia \and Geneva Observatory, 1290 Sauverny, Switzerland}
\date{Received / Accepted}
\abstract{We compute the evolution of different abundance
ratios in the Milky Way (MW) for two different sets of stellar yields. In
one of them stellar rotation is taken into account and 
we investigate its effects on the chemical evolution model predictions. 
Moreover, we show that some abundance ratios 
offer an important tool to investigate the halo-disk discontinuity. 
For the first time it is shown that the effect of a halt in the star
formation between the halo/thick disk and thin disk phases, already
suggested from studies based both on Fe/O vs. O/H and Fe/Mg vs. Mg/H,
should also be seen in a C/O versus O/H plot {\it if C is produced mainly
by low- and intermediate-mass stars} (LIMS). 
The idea that C originates mainly from LIMS is suggested by the flat 
behavior of the [C/Fe] ratio as a function of metallicity, from 
[Fe/H]$\sim -$2.2 to solar, and by the fact that very recent C/O 
measurements for stars in the MW halo and disk 
seem to show a discontinuity around log(O/H)+12$\sim$8.4.
Finally, a more gentle increase of N abundance with metallicity (or time), 
relative to models adopting the yields of van den Hoek \& Groenewegen (1997),
is predicted by using the stellar yields of Meynet \& Maeder (2002 - 
which include stellar rotation but not hot-bottom 
burning) for intermediate mass stars.
This fact has some implications for the timescales of 
N enrichment and thus for the interpretation of the nature of 
Damped Lyman Alpha Systems.

}

\begin{document}
\maketitle
\section{Introduction}
In the past years new observations have suggested that
the Milky Way (MW) formed in a more complex way 
than the pure dissipational collapse suggested originally by
Eggen, Lynden-Bell \& Sandage (1962, ELS). Their model assumed 
a continuous evolutionary transition in the formation of the halo, thick disk
and the thin disk. Apart from dynamical arguments (Wyse \& Gilmore 1992), 
the chemical
properties of the MW disk are convincingly showing us that its formation
was neither smooth nor continuous. As an example, 
Gratton et al. (1996, 2000) presented a compilation of stars for which
both the kinematics and the [Fe/O] abundance ratios were known. They
observed for the first time a discontinuity in the [Fe/O] vs. [O/H]
plot which was interpreted as a halt in the star formation before
the thin disk formation. In particular, they observed that around 
[O/H]$=-$0.2 dex (in solar scale) the oxygen abundance remains constant
whereas the [Fe/O] ratio keeps increasing. This is interpreted as a break
in the star formation since in this case oxygen would not be produced
whereas Fe would continue to originate from long-living systems giving
rise to type Ia SNe. In this scenario no stars would have been formed during
the gap and this implies that a gap should also be seen in an
[Fe/$\alpha$] vs. [$\alpha$/H] diagram. The same behavior
was found by Fuhrmann (1998) for [Fe/Mg] vs. [Mg/H]. However, this 
should still be confirmed by larger data samples.

Chiappini et al. (1997) have shown that a two-infall model for the 
formation of the MW, 
where the halo (and part of the thick disk)
formed on a short timescale whereas the thin-disk formed on a much
longer one, can explain
not only the above observations but also provides a good fit to the 
G-dwarf metallicity distribution in the solar vicinity
(see also Kotoneva et al. 2002). In the present work we argue that, if we 
believe that there was such a halt in the star formation rate (SFR) and if the 
carbon enrichment of the interstellar medium (ISM) is mainly due to low and 
intermediate mass stars (LIMS) and hence is produced
on long timescales as iron, 
the same kind of discontinuity observed for Fe/O and Fe/Mg should be 
seen in a log(C/O) vs. log(O/H) diagram (Chiappini et al. 2003 - CRM2003).

However, the origin of C (and
N) we see today in the ISM is still
an open problem. 
In particular, while Henry et al. (2000) favor the hypothesis 
that most of the carbon we observe today in the ISM comes 
from massive stars, CRM2003
suggest that most of the carbon comes from LIMS. 
These conclusions are very dependent on the adopted stellar yields.
Many are the processes involved in the computation of the stellar yields
of C and N and there are still many uncertainties present in these 
calculations (see Meynet \& Maeder 2002 - MM). MM have
shown that stellar rotation can affect the predictions of the stellar yields
especially for He, C, N and O.
Chemical evolution models can thus be used to test and constrain 
the stellar yields, as it will be shown in the next sections. In
particular, we will evaluate the impact of the new stellar yields
of MM on important open questions related to the C, N and He enrichment
in galaxies. Moreover, we discuss how the abundance ratios depend on the adopted stellar
yields and also how they can be used to infer the star formation history
of our Galaxy and the nature of Damped Lyman Alpha systems (DLAs).

\section{Stellar Yields}

In this work we compare the following two different models:
a) a model computed with van den Hoek \& Groenewegen (1997 - vdHG) 
yields for LIMS and 
Woosley \& Weaver (1995 - WW) yields for massive stars (model A) and 
b) a model computed with the recent
published stellar
yields of Meynet \& Maeder (2002 - MM), which take into account
the effects of rotation on stellar evolution, for the whole range of 
stellar masses (model B). In model A the carbon yields are
multiplied by a factor of three in the 40-100 M$_{\odot}$ mass range\footnote
{WW computed the stellar yields for stars in the 11 to 40M$_{\odot}$
mass range. For masses larger than 40M$_{\odot}$ we extrapolated their 
stellar yields, as shown by the triangles in Figs.~1 and 3.
In particular, for carbon, our extrapolated value at 70M$_{\odot}$ 
was chosen to match the value computed for this mass by Nomoto et al.
(1997 - these latter are essentially the same as in Thielemann et al. 1996, 
but for an enlarged grid of masses - see open triangle and open
circle in Fig.~1 for a 70M$_{\odot}$). However, as concluded in CMR2003, 
although the yields of Thielemann et al. (1996) 
ensure a good agreement between
data and model predictions for many abundance ratios,
for carbon it was necessary
to increase the stellar yields, for $m>$40M$_{\odot}$, by a factor
of three. In fact, Thielemann et al. (1996) do not account for
mass loss which, during the WC phase, is responsible for ejecting
helium and carbon into the interstellar medium.
Given this fact, in the present work we also increased
the WW yields (which also do not account for mass loss) by the same factor 
(see filled triangles in the upper panel of Fig.~1).}

Three important facts should be noted with respect to the different
sets of yields:

\begin{itemize}
\item
The vdHG yields we adopt here are for the case of $\eta_{\rm AGB}$ (the mass
loss parameter during AGB phase) varying with
metallicity ($\eta_{\rm AGB}$ =~1 and 2 for Z =~0.001 and 0.004 respectively,
and $\eta_{\rm AGB}$ =~4 for Z=~0.008, 0.02 and 0.04). We notice that a lower
value for $\eta_{\rm AGB}$ implies a larger yield of carbon. This is because 
a lower mass loss leads to a longer lifetime and hence more thermal pulses.
As a consequence, more C is dredged up to the stellar surface.
An example of the impact of the mass loss parameter (which is one of
the main uncertainties in the computation of vdHG yields, together with
the efficiency of the hot bottom burning - HBB) is shown in Fig.~6 (upper left panel). 
The dashed line shows a model
computed with vdHG yields for Z=~0.001 and
$\eta_{\rm AGB}$ =~1 for all metallicities. In this
case a high log(C/O) ratio is predicted at the present time (see a detailed
discussion in the Sect.~4).
\item
MM use a different approach from vdHG. While the 
latter authors computed their yields by means of synthetic 
AGB models, MM yields were obtained from self-consistent
complete stellar models (without any fine tuning of parameters
related to the so called HBB).
MM were able to show that rotation opens a 
new alternative for primary N production in intermediate mass stars,
in addition to the classical HBB scenario. Moreover they also
predict some primary N production in massive stars.
However, for the intermediate mass star models MM calculations
stop at the beginning of the thermal-pulse AGB phase (TP-AGB).
The third dredge-up (and the HBB) would occur in more evolved stages
and thus are not included in the MM results.
Only, in the case of models with Z$=$10$^{-5}$ they do obtain the 
third dredge-up in the sense that the stellar surface becomes enriched 
both in H and He-burning products. Forthcoming results (Meynet, private
communication) will include the more evolved phases of stellar
evolution thus predicting not only the 3rd dredge up contribution
to $^{12}$C but also obtaining the HBB effect as a natural 
consequence of the stellar evolution itself, without any parametrization. 
In fact, HBB may still appear and thus add its contribution to the 
synthesis of primary N.
\item 
WW models included the explosive nucleosynthesis,
but did not include rotation/mass loss. Their tabulated yields, 
differently from the ones given in MM (or Thielemann et al. 1996 - TNH) 
include not only the processed material (i.e., the new elements
produced and released by a given star of mass $m$), but also the unprocessed
material of the stellar envelope. This last quantity has to be subtracted
from their tables if one wants to use their stellar yields in a consistent
way.

\end{itemize}

In summary, one has to keep in mind that the stellar yields of MM offer
a new alternative for the primary N production in intermediate mass stars
and in massive stars. However, their
results do not include the third dredge-up and hot-bottom burning and thus
their yields of C and N for the intermediate mass range 
should be taken as lower limits. On the other
hand, the vdHG yields, which include both HBB and 3rd dredge-up, 
depend strongly on the $\eta_{\rm AGB}$ and HBB efficiency parameters
adopted in their synthetic models and are thus very uncertain. 

Although MM did not formally include the third dredge-up, we 
think it is worth studying the effects of their new yields on 
chemical evolution models for the following reasons:
a) MM yields for nitrogen at 
low metallicity results from a new process whose importance 
for chemical evolution has to be studied.
In absence of a real quantitative assessment of the 
importance of the HBB it appears to us interesting to 
study the importance of this new process,
which produce ``non-parametric'' yields, independently of HBB
and b) this is particularly justified in view of the fact 
that this new process give primary N yields at low 
metallicity not very different from those obtained 
from parametric studies as the one of van den Hoek 
and Groenewegen. This questions the importance
of the HBB. Only by studying the effects separately 
it will be possible to understand the different 
consequences of the two processes.

\subsection{The yields adopted in this work}

We are interested to study the effects of
stellar yields which account for stellar rotation 
on chemical evolution models.
In model B we adopted the yields of MM for stellar rotational velocities
of 300 km/s (MM published only the tables for Z$=$10$^{-5}$ and 0.004. Their
computed values for Z=~0.020 and V$_{\rm rot}=$300km/s are shown in Table 1.
These yields have been deduced from the models of Meynet \& Maeder 2000). 
MM provide also calculations for larger velocities
(400 km/s) but only for some specific masses and metallicities, namely:
for N at Z=~0.004 (for a mass of 20 M$_{\odot}$) and for
C, N and O for Z=~10$^{-5}$ (for stars of 9 and 20 M$_{\odot}$). In the
case of Z=~0.004 the stellar yields computed with rotational velocities
of 300 km/s or 400 km/s are the same
(see their Table 5). For their lowest metallicity
calculations, the yields computed with a velocity of 400 km/s are similar to those computed
with 300 km/s for He, C and O. Only for N there is a difference, namely the 
yield computed with V$_{\rm rot}=$400 km/s is 
a factor of two larger than the one computed with V$_{\rm rot}$=~300 km/s for a 20 M$_{\odot}$ and 
a factor of 3 larger for a 9 M$_{\odot}$.

\begin{table}
\caption{\small{Meynet \& Maeder yields for Z=~0.020 and V$_{\rm rot} =$300km/s}}
\small
\begin{tabular}{lllll}
\hline
Initial Mass & m p$_{^4He}$ & m p$_{^{12}C}$ & m p$_{^{14}N}$ &  m p$_{^{12}O}$ \\
\hline
120 &    31.347 &  7.114 &  0.80 &   7.00  \\
60  &    15.269 &  2.161 &  0.35 &   3.48  \\
40  &    5.764  &  1.732 &  0.13 &   3.80  \\
25  &    2.525  &  0.310 &  0.12 &   2.88  \\
20  &    2.621  &  0.178 &  0.11 &   1.41  \\
15  &    2.201  &  0.099 &  0.09 &   0.41  \\
12  &    1.761  &  0.014 &  0.07 &   0.07  \\
9   &    1.056  &  0.055 &  0.03 &   0.017 \\
7   &    0.744  &  0.037 &  0.02 &   0.011 \\
5   &    0.433  &  0.018 & 0.014 &   0.005 \\  
3   &    0.121  & .00031 & .0065 &  -0.0014 \\
2   &    0.067  & .0001  & .0025 &  -0.0014 \\ 
\hline
\end{tabular}
\label{table1}
\end{table}

\begin{figure}
\centering
\includegraphics[width=8cm,angle=0]{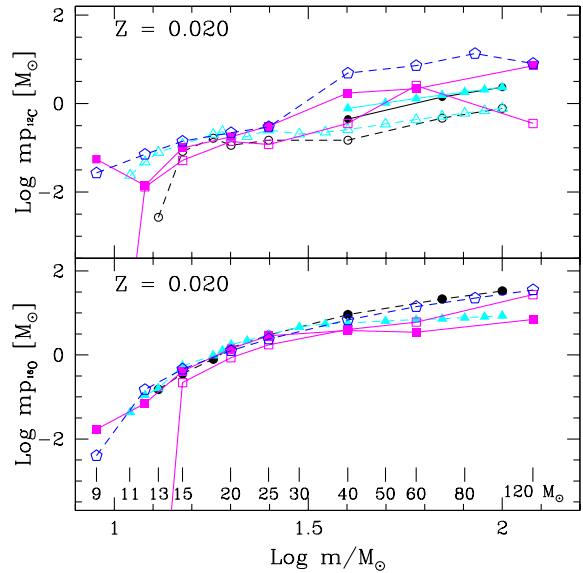}
\caption{Different stellar yields for $^{12}$C (upper panel) and $^{16}$O
(lower panel) as functions of the
initial stellar mass for massive stars, at solar metallicity: MM
for V$_{\rm rot} =$0 km/s (open squares) and V$_{\rm rot} =$300 km/s (filled squares); Thielemann
et al (1996) (circles);  WW (triangles); Maeder (1992) (open
pentagons). 
In the upper panel,
the filled circles and filled triangles show the carbon yields of Thielemann
et al. (1996) and WW, respectively, multiplied by a factor
of 3 in the 40-100 M$_{\odot}$ mass range (see text).}
\end{figure}

In Fig.~1 we compare the stellar yields for carbon and oxygen as computed
by different authors, for the solar metallicity. For comparison we also show 
the yields of Maeder (1992) computed with strong mass loss and no rotation.
It is clear from the figure (upper panel)
that the yields of Maeder (1992 - open pentagons) for carbon are 
larger than those of MM even for the case of a large rotational 
velocity (filled squares).  This difference arises mainly from the fact that in MM the mass loss rates are lower than those used in Maeder (1992). In particular, in MM the reduced mass loss rates accounting for the effects of clumping have been taken into account during the Wolf-Rayet phase.
From Fig.~1, one sees that the WW and 
TNH yields multiplied by a factor of 3 in the 40-100 M$_{\odot}$ mass range are similar
to the new calculations of MM (see CMR2003). 
For oxygen (lower panel) the figure shows that the yields of 
Maeder (1992 - open pentagons) are larger than those of WW (triangles) or 
the ones given by MM. There is a good agreement among all the oxygen yields 
in the 15-25 M$_{\odot}$ mass range.

\begin{figure}
\centering
\includegraphics[width=8cm,angle=0]{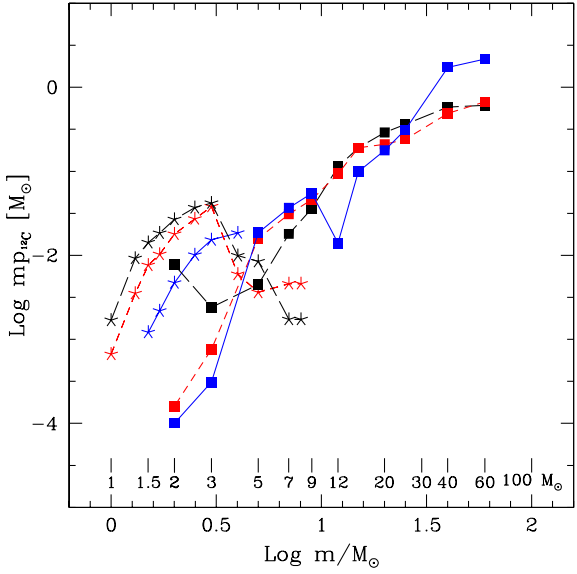}
\caption{Comparison of MM (squares) and vdHG (asterisks) yields of $^{12}$C for 3 different
values of metallicities (solid lines: solar, dashed line: Z=~0.004 
and long dashed line: Z=~0.001 for vdHG and Z=~0.00001 for MM), 
for masses up to 8M$_{\odot}$. Above 9M$_{\odot}$
we show the yields of MM for different metallicities.}
\end{figure}

In Fig.~2 we compare the predictions of MM (squares) for LIMS with those of vdHG (asterisks)
for $^{12}$C, for different metallicities. It is clear from the figure that
for all metallicities, vdHG produce more carbon than MM. As discussed before
this is mainly due to the fact that MM did not include the 3rd dredge up in their calculations 
and hence their yields for carbon, in this mass range, should be seen as lower limits.
For Z=~10$^{-5}$ (squares connected by a long-dashed curve), MM do include the 3rd 
dredge up and this is why in this case more carbon is obtained. For the most massive stars 
the yields of MM predict more carbon at solar
metallicities than in the two low metallicity cases. 
This is an effect of mass loss already present in Maeder (1992) 
calculations but which is less important in the new calculations of MM.

\begin{figure}
\centering
\includegraphics[width=8cm,angle=0]{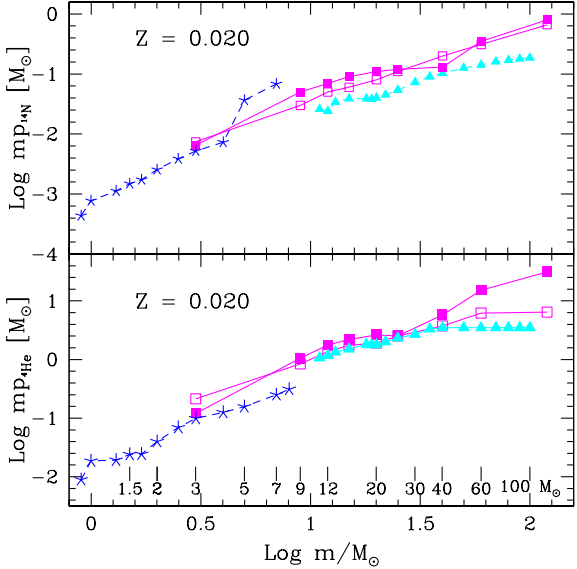}
\caption{Different stellar yields for $^{14}$N (upper panel) and $^{4}$He
(lower panel),
for the whole stellar mass range. The symbols are as in Fig.~1. The 
asterisks represent the yields of van den Hoek and Groenewegen (1997).}
\end{figure}

In Fig.~3 we compare the different stellar yields computed for
$^{14}$N (upper panel) and $^{4}$He
(lower panel) for the whole mass range,
for solar metallicities. 
For massive stars, mass loss by stellar winds, which is taken into account in the computation of MM,
lead to larger yields of N and He relative to the work of WW (triangles).
In the intermediate-mass range (5 to 8 M$_{\odot}$)
the nitrogen yields of vdHG (asterisks) are larger than the ones of MM and this is due to the
contribution of the HBB. For the He helium production, MM yields
with rotation predict more helium than WW for massive stars, and also
more than vdHG for intermediate mass stars. The consequence of the
larger helium production when adopting the yields of MM on the $\Delta Y/\Delta Z$
ratio will be shown in Sect.~6.

\begin{figure}
\centering
\includegraphics[width=8cm,angle=0]{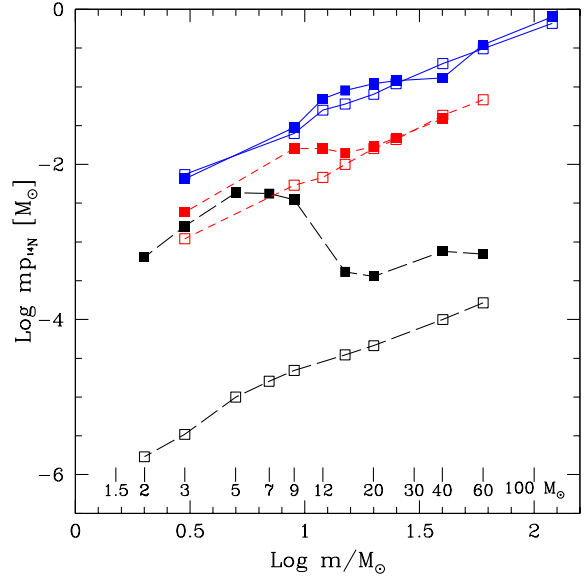}
\caption{MM stellar yields for $^{14}$N (upper panel),
for the whole stellar mass range, for different metallicities. 
The symbols and lines are as in Fig.~2. The yields of MM for stellar
models were rotation is not taken into account are also plotted (open squares).}
\end{figure}

In Fig.~4 we plot the yields of MM (squares) 
for $^{14}$N, for different metallicities. Open symbols refer to
stellar models were rotation is not taken into account, whereas filled
squares refer to yields obtained from stellar models with V$_{\rm rot}$=300km/s.
It is clear from the figure that rotation does not change much the nitrogen
yields at solar metallicities (squares connected by a solid line), 
while it makes some difference at Z=0.004 (squares
connected by short-dashed line). However, an important difference
is seen at low metallicities (in this case Z=10$^{-5}$ - squares connected by
long-dashed line). The yields with rotation
of MM are similar to the ones of vdHG (for the case of 
$\eta_{\rm AGB}$ variable with metallicity), 
for all metallicities, for $m<$4M$_{\odot}$, whereas,
for masses between 4 and 8M$_{\odot}$, MM yields are substantially smaller.
This is due to the strong HBB assumed in vdHG calculations.

\section{Chemical Evolution Model of the Milky Way}

In the next sections we show the results of chemical evolution models
for the MW computed with the two different sets of yields discussed
before. The model is that of CMR2003 (see also Chiappini et al. 2001, 
where a detailed description can be found). As explained in Sec.1, 
the fundamental idea of this model is that
the formation of the MW occurred in two different infall episodes,
one forming the halo and part of the thick disk on a relative short
timescale and another one forming
the thin-disk on a longer timescale. In this model a threshold gas density
is assumed and, as a consequence, the star formation rate becomes zero
every time the gas density drops below the threshold value. 
The two-infall approach, combined with such a threshold, leads
to a gap in the star formation before the formation of the thin-disk. During 
the ``gap'' in the star formation only elements produced by 
type Ia SNe and LIMS, born before the ``gap'', are restored into the ISM. As
a consequence this model predicts an increase in the abundance
ratios of elements restored on long-timescales over $\alpha$-elements 
(produced basically by massive short-lived stars) around
a metallicity of [Fe/H]$\sim -$0.6 dex (which corresponds to the time 
of the halt in the SFR which we predict to be around 10 Gyrs ago 
- see Chiappini et al. 1997 for details). 

\section{Carbon enrichment}

\subsection{The [C/Fe] vs. [Fe/H] diagram}

\begin{figure}
\centering
\includegraphics[width=8cm,angle=0]{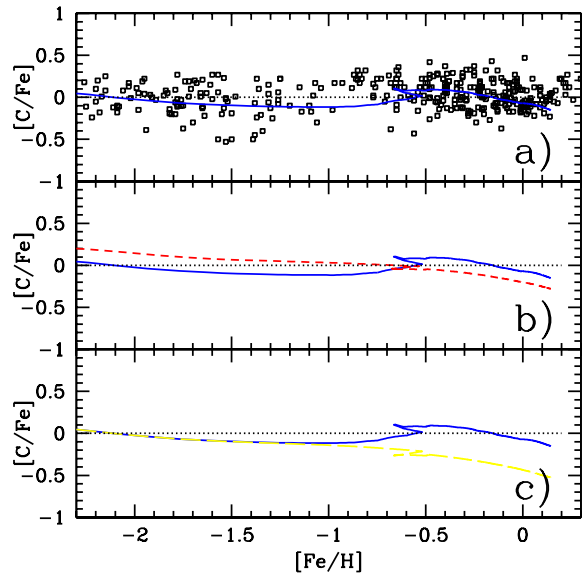}
\caption{[C/Fe] vs. [Fe/H]. The dotted line in all panels markes
the solar ratio. Panel a) shows a model computed with 
vdHG + WW yields (model A - solid line) compared
to abundance data from the compilation of Chiappini et al. (1999).
Both the model predictions and the abundance data show an almost
solar [C/Fe] ratio (down to [Fe/H]$\simeq -$2). In panel b) we compare
the model A shown in a) with a model computed with MM yields (model B - short dashed line). 
Panel c) shows again the model A (solid line) compared with a similar model
where the contribution from LIMS is suppressed (long-dashed line).}
\end{figure}

In Fig.~5 we show the variation of the [C/Fe] ratio as 
a function of metallicity
for stars in the MW compared with the predictions of model A 
(vdHG + WW yields - solid line).
The observed solar [C/Fe] ratio from [Fe/H] $\sim -$2 up [Fe/H]$=$0 
indicates that C and Fe are produced on similar
timescales (CRM2003). This implies that the main producers of C cannot be massive
stars unless the stellar yields vary strongly with metallicity. 
However, on the basis of the most up to date stellar models this latter
alternative is unlikely and hence our conclusion is that the C 
we see today in the ISM comes mainly from
LIMS, as originally suggested by Tinsley (1979).
For metallicities below [Fe/H] $\sim -$2, type Ia and 
intermediate-mass stars did not have yet time to contribute 
to the enrichment and the behavior of [C/Fe] can be substantially 
different as it will depend essentially on the variation of 
the ejected masses of C/Fe from massive stars.

In Fig.~5b we compare the [C/Fe] predicted by model A (solid line) and
model B (computed with MM stellar yields - short-dashed line). 
Model B results are good for low metallicities,
but predicts too few carbon during the thin disk phase 
([Fe/H] $>-$0.8). The main reason for that is the absence of the 3rd 
dredge up in MM models for metallicities larger than Z$=$10$^{-5}$.
In fact, as shown in Fig.~5c, a similar 
behavior is seen if in model A
we ``turn off'' the contribution to C from LIMS (long-dashed line curve).

\subsection{The log(C/O) vs. log(O/H) diagram}

\begin{figure*}
\centering
\includegraphics[width=12cm,angle=0]{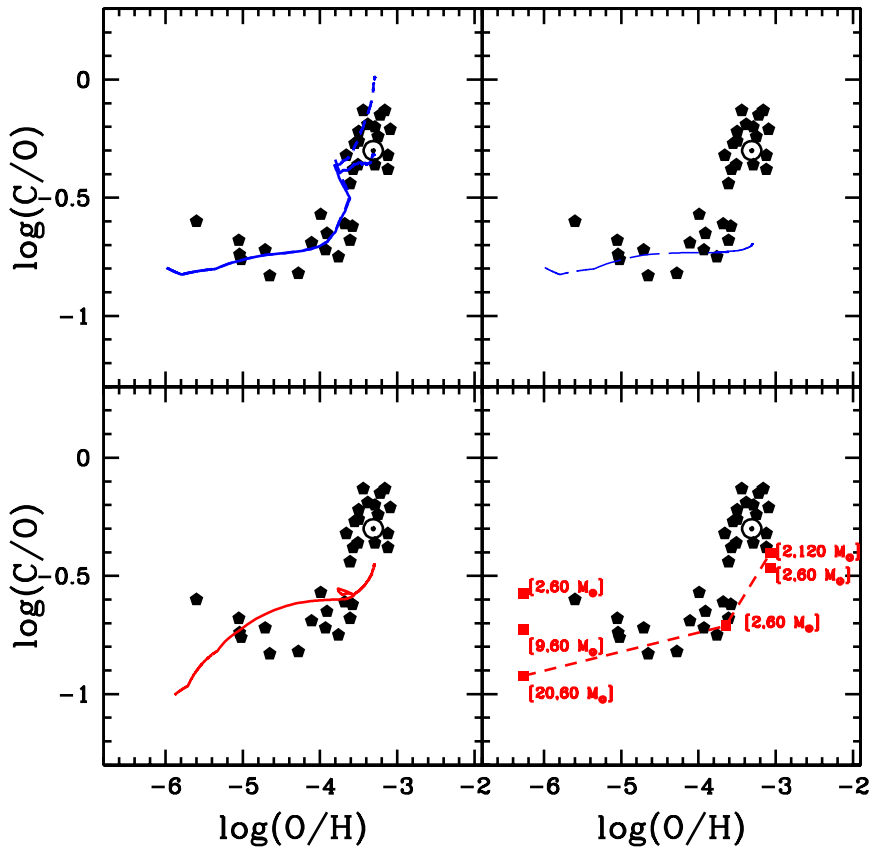}
\caption{Model predictions for the galactic evolution of the
C/O ratio as a function of O/H compared with data
obtained by Nissen (2003). The abundance data
include only stars in the Milky Way (halo, thick disk and thin
disk stars) whose abundances are representative of the ISM
from which they formed. 
The solar value from Allende-Prieto et al. (2001, 2002)
is also shown. In the left upper panel a chemical evolution model for the MW
computed with vdHG + WW yields (model A) is shown (solid line).
The dashed line shows the same model computed with stellar yields for LIMS 
given in the table for Z=~0.001 from vdHG with $\eta_{\rm AGB}$ =~1 (an upper limit
for the C production in LIMS). In the right upper panel we show again  
model A but with no contribution from LIMS. In the lower left panel
model B prediction is shown (computed with stellar yields including
rotation effects from Meynet \& Maeder 2002). In the lower right panel
we plot directly
the integrated yields (for different ranges of masses) 
as at different metallicities (without the use of a chemical
evolution model). This is 
a simplistic approach that implicitly assumes instantaneous recycling approximation
and thus neglects the effects due to stellar lifetimes.}
\label{sample-figure}
\end{figure*}

\subsubsection{Abundance data}

Several papers devoted to the discussion of the log(C/O) vs.
log(O/H) relation and its interpretation, consider the abundance
data of different galaxies together with the abundance data
of the MW (e.g. Carigi 2000, Henry et al. 2000). As
discussed in CRM2003 the abundance data for the 
MW contain an important temporal information in the 
sense that low metallicity main sequence stars represent the composition of the
ISM at the time of their formation. This is not the case 
if we plot together the present time abundances in HII regions in other 
galaxies. In addition, each galaxy has suffered a particular enrichment
history. Given this fact we decide to plot here
only the stellar abundances of C and O observed in the MW. Moreover,
since our last paper (CRM2003) a new homogeneous data sample was
published by Nissen (2003) which include halo, thick disk and thin 
disk stars. This data set shows a very important
new feature (see Fig.~6): it seems to indicate a 
discontinuity between the halo and
thin disk stars. Moreover, this discontinuity  seems to happen exactly
at the same metallicity where Gratton et al. (1996)
and Fuhrmann (1998) observed a lack of stars in Fe/O and Fe/Mg plots, respectively. 
More data are necessary to confirm this possible gap in the stars.

\subsubsection{Model predictions}

In Fig.~6 (upper left panel) we compare the data mentioned before 
with our model A prediction (solid line). 
As it can be seen, a model in which a halt in the
star formation is predicted and including stellar yields where an 
important amount of C comes from LIMS, produces a 
discontinuity in the log(C/O) vs. log(O/H) diagram, 
similar to what is observed. This would not be the
case if C were mainly produced from massive stars. In fact, in the same figure (upper right panel)
we show model A but with no contribution from LIMS to carbon.
In this case, as expected, we obtain a flat log(C/O) 
along the whole metallicity range.

Fig.~6 (lower left panel) shows the prediction of model B (computed with
MM yields). Two things can be noted: a) the present log(C/O) is too low and
b) the discontinuity discussed before is not present anymore. In fact, since in this case
the LIMS do not produce much C (due to the lack of the 3rd dredge-up), we see
essentially what would be expected for the ratio of two elements produced by 
massive stars namely, during the star formation ``gap'' both 
elements stop being produced,
and the only effect seen is the decrease in the O/H ratio due to 
the second infall that forms the disk (assumed to be primordial) 
which produces
the ``looping'' seen in the figure.
In the low right panel we show the effect of assuming instantaneous 
recycling approximation (I.R.A.) when trying to interpret
the log(C/O) abundance ratio. As C is restored into the ISM on longer timescales
than O, the I.R.A. is not correct in this case. We illustrate this point
by plotting the integrated yields of MM computed for different metallicities and integrated
over different mass ranges (indicated in the figure in brackets), 
assuming a Scalo IMF, as in the chemical evolution models presented here. 
As we can see, the log(C/O) at solar metallicity, obtained by integrating the stellar
yields, is slightly larger than what is obtained by using the same stellar yields in a chemical
evolution model which takes into account the stellar lifetimes. In fact,
the final C/O value achieved by model B (in the lower left panel) 
is lower than that obtained when assuming I.R.A. (right lower panel). 
Because of the relaxation of I.R.A. in our models, the
low mass stars formed out of gas enriched in oxygen relative to iron or carbon, 
at early times, restore
their pristine gas with C/O ratios characteristic of the halo phase, thus lowering
the present C/O ratio.

In summary, the contribution
of LIMS to the abundance of C is important. The opposite conclusion 
was suggested by some authors and 
arised for two reasons: a) the difficulty of reproducing the C/O ratio 
versus O/H at large O abundances and b) the yields of Maeder (1992), where 
a strong mass loss, operating in stars of high metal content, allows the most 
massive stars to loose more and more carbon as a function of stellar mass 
and metallicity. This solution creates some difficulty then in explaining 
the [C/Fe] vs. [Fe/H] running flat for all
metallicities (as shown in CMR2003). Moreover, as shown in this section, 
the new yields of MM do not predict
anymore a rise in the C/O ratio as a function of oxygen of the same 
magnitude as it would be obtained with Maeder (1992) yields.
Our results show that the solar C/O value can be explained by models 
which predict a break in the star formation rate before the thin disk formation together
with stellar yields for intermediate mass stars including the 3rd dredge up.

\section{The N enrichment}

\subsection{The Milky Way}

Ideally the best way to ascertain the nature of nitrogen would be to 
look at the N abundances in the stars in the MW since they represent a true
evolutionary sequence, where the stars with lower metallicity are the oldest
ones. Unfortunately, the N abundances for stars in the MW
are still uncertain, especially at low metallicities (see Nissen 2003).
Current data show a [N/Fe] ratio which is below solar at low metallicities and
then increases and flattens in disk stars (see CRM2003). 
This behavior agrees with
the assumption that primary N production in massive stars
is small.
Here we show that this is still true even when the yields of MM, which predict
some primary N contribution from massive stars, are adopted. 

\begin{figure}
\centering
\includegraphics[width=8cm,angle=0]{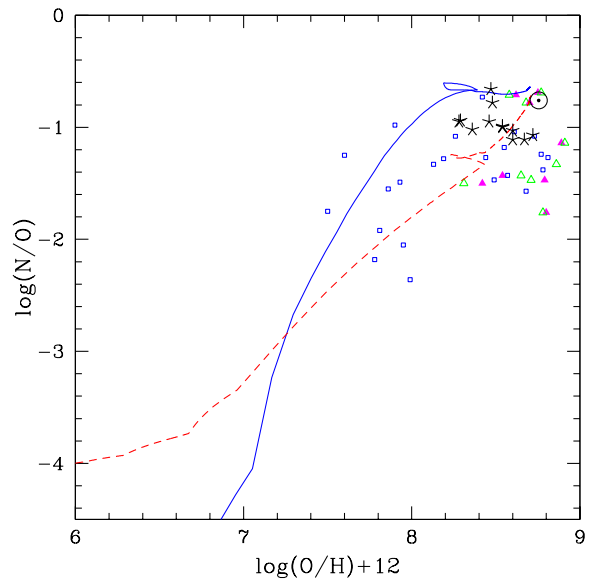}
\caption{Diagram showing log(N/O) vs. log(O/H)+12 for the Milky Way (model A: solid
curve, and model B: dashed curve; see CRM2003
for a description of the abundance data)}
\end{figure}

\begin{figure}
\centering
\includegraphics[width=8cm,angle=0]{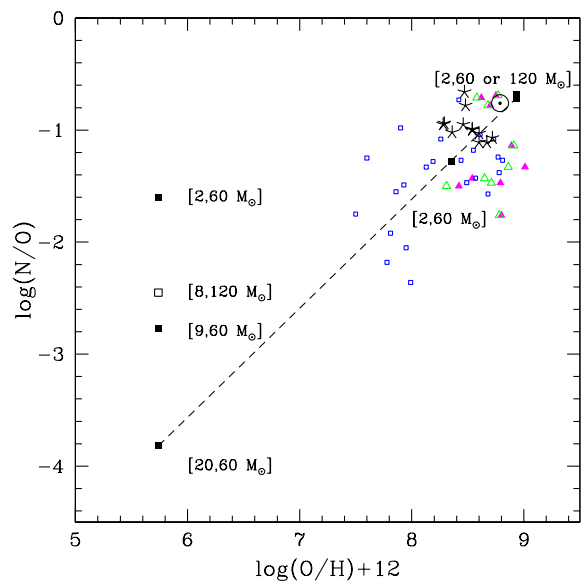}
\caption{Diagram showing log(N/O) vs. log(O/H)+12 obtained by integrating the MM yields
in different mass ranges as indicated in brackets. As it can be seen by comparing with
Fig.~7 the main stars contributing at metallicities of the order of log(O/H)+12 $\sim$ 5.8
are those above $\sim$20M$_{\odot}$.The empty square refers to the case with rotational velocity of 400 km $sec^{-1}$.}
\end{figure}

In Fig.~7 we show the predictions 
of models A (solid line) and B (dashed line) for log(N/O) vs. log(O/H)+12 for the solar vicinity.
The data points are from a compilation by CRM2003. 
It can be seen that the primary N production in massive stars predicted by model B,
with the MM yields, is small and gives rise to a sort of ``plateau'' at
log(N/O) $\sim -$4 and log(O/H)+12 $\leq$ 7. In fact, almost
the same value is obtained by MM when the stellar yields corresponding to
log(O/H)+12 =~5.74 and
velocities of 300km/s, are integrated in the 20-120 M$_{\odot}$ mass range, which
should be the stars contributing to the ISM enrichment at such 
low metallicities in the case of the MW (see Fig.~8). 
This is the first result, namely that the
primary N contribution from massive stars predicted by MM is small.
A second result is that the increase of N in intermediate mass
stars in model B is more gentle than
in model A where the HBB contributes a substantial amount primary N.
Models A and B seem to agree with the available data for the MW
but more data are clearly needed, especially at low metallicities.

\begin{figure}
\centering
\includegraphics[width=8cm,angle=0]{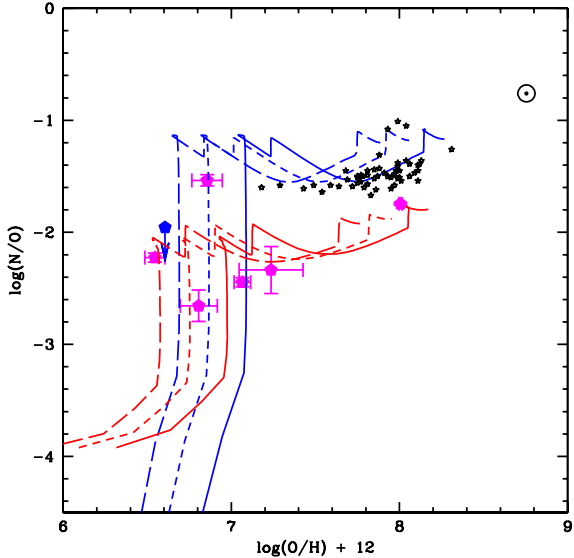}
\caption{The same diagram as in Fig.~6 but now for  
dwarf galaxies of Izotov \& Thuan (1999) and DLAs from Centurion et al. (2003a)
(filled pentagons  - with a new determination from Centurion et al. 2003b).
Model predictions are explained in the text. For the lower family of curves
an abundance ratio log(N/O) $\sim -$2.5 is attained after 53 Myr from the start
of the star formation, while for the upper curves log(N/O)$=-$1.8 after 53Myr.}
\end{figure}

The above results show that a plateau,
due to primary N production in massive stars as suggested by MM yields, occurs at a log(N/O)
ratio which is far below (log(N/O)$\sim-$4.0 dex when the yields of MM for 
V$_{\rm rot}$=~300km/s are adopted) 
the ``plateau'' seen either in blue compact galaxies (BCGs) at a log(N/O)$\sim-$1.6 dex  
(Izotov \& Thuan 1999) or in DLAs at a log(N/O)$\sim-$2.2 dex. Only when very
high rotation velocities are considered, the primary N production by massive stars may become significant (see Fig.~8 the empty square which corresponds to V$_{\rm rot}$=~400km/s).

Izotov and Thuan (1999) have claimed that 
BCGs show a flat log(N/O) plateau at 
low metallicities and suggested it to represent the minimum N/O value due to the 
primary N production in massive stars. As discussed in CRM2003 we think this 
interpretation is not correct since in the case of BCGs we are comparing 
different objects at present time and the observed N/O ratios are therefore the 
products of 13 Gyrs of evolution. For BCGs this diagram
is not the same as for the MW stars, in the sense that the x-axis (log(O/H)+12)
cannot be interpreted as a time axis. Moreover, if that was the case a plateau
around log(N/O)$\sim-$1.6 should also be seen for the MW stars. Current data for the MW
do not support this idea (CRM2003). 
However, the models shown in Fig.~7 apply only to the 
solar vicinity. In what follows we would like to better assess the impact of the 
MM yields on the interpretation of BCGs and DLAs abundance data by computing
chemical evolution models which apply to these systems.

\subsection{Blue compact galaxies and DLAs}

Blue compact galaxies are best reproduced by short and moderately
intense bursts of star formation followed by longer 
quiescent periods (Matteucci \& Chiosi 1983).
As shown in CRM2003, the large spread observed both in the N/O versus O/H
and C/O versus O/H diagrams for oxygen abundances larger than 7.6 can be 
explained as arising from different chemical evolution histories of different
galaxies (i.e. different star formation efficiencies, different burst ages
and different burst durations). The nature of DLAs is still
a matter of debate but their chemical characteristics seem to be in agreement 
with the idea that they represent either the progenitors of the current dwarf
galaxies or the outer parts of disk galaxies (see Prantzos \& Boissier 2000,
Lanfranchi \& Matteucci, 2003 and Calura et al. 2003).
In particular, Lanfranchi \& Matteucci (2003) suggest that some of the DLAs
chemical properties can be well fitted by models for BCGs, with 
roughly 4 bursts of star formation, with star formation efficiencies in the 
range of 0.1 - 0.9 Gyr$^{-1}$. This kind of models
are able to fit the [$\alpha$/Fe] observed in DLAs but still predict too much
N when the yields of vdHG are adopted. 

In Fig.~9 we show the same kind of diagram as in Fig.~7 but now we plot the
DLA abundance data (from
Centurion et al. 2003a,b - pentagons) for objects where oxygen was 
measured or a limit was given. Also shown are the BCGs observed by 
Izotov \& Thuan (1999) (small symbols). 
The curves show ``bursting'' models computed with four
bursts and a Salpeter IMF\footnote{In 
CRM2003 we suggest that to be able to fit the N/O ratio observed in BCGs
a flatter IMF was necessary, but Lanfranchi \& Matteucci (2003) concluded
that such models would predict [O/Fe] ratios larger than the observed ones.}.
In these models, similar to Lanfranchi \& Matteucci (2003), we assumed 
4 bursts at t=~1, 10, 13 and 13.98 Gyrs with a duration
of 0.02, 0.01, 0.2 and 0.02 Gyrs, respectively. 
The ``family'' of upper curves shows models computed with different
star formation efficiencies (0.1, 0.2 and 0.3 Gyr$^{-1}$ shown as long-dashed, 
short-dashed and solid lines, respectively) and with the same
nucleosynthesis prescriptions of model A (i.e. vdHG + WW), whereas
the lower curves show the same models computed with
the yields of MM. In this case the DLAs N abundance
can be well explained. 

In fact,
given the more gentle increase of the N when the yields of MM are adopted (due to
less N from LIMS) it is possible
to obtain a low log(N/O) without predicting a too high [$\alpha$/Fe]
enhancement, as required by observations in DLAs. 
When the MM yields are adopted, a log(N/O) ratio of $-$2.5
is achieved after roughly 53 Myrs (see Fig.~10). This means that intermediate
mass stars are already contributing 
for the N enrichment even in DLAs which show a mean log(N/O)$\sim-$2.2 dex 
(we recall that the lifetime of an 8M$_{\odot}$ is of the order
of 30Myr). Models with vdHG yields can also explain the DLA abundances, but
in this case DLAs would be very young systems and this may happen
as an odd coincidence. 

\begin{figure}
\centering
\includegraphics[width=8cm,angle=0]{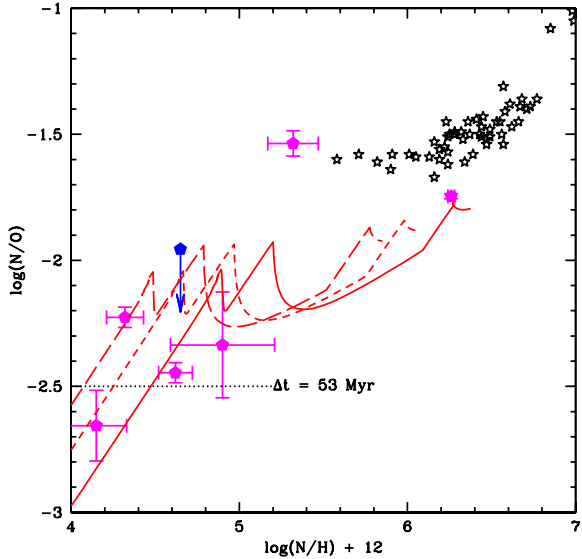}
\caption{Log(N/O) vs. log(N/H)+12 diagram. 
Dwarf galaxies are from of Izotov \& Thuan (1999 - black dots); 
DLAs from Centurion et al. (2003a,b) (filled hexagons).
The models are the same shown in Fig.~9 (lower family of curves, 
with MM yields). 
The $\Delta t =~53$ Myr indicates the time elapsed
from the beginning of star formation and the dotted line represents an isochrone. In other words, the objects on that line have the same age but different star formation efficiency.} 
\end{figure}

\begin{figure}
\centering
\includegraphics[width=8cm,angle=0]{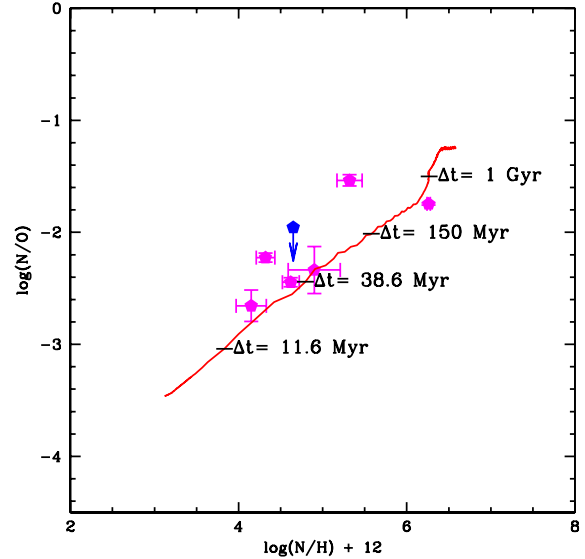}
\caption{Log(N/O) vs. log(N/H)+12 diagram. 
DLAs from Centurion et al. (2003a,b) (filled hexagons).
Model prediction for MW (with MM yields) at an outer region - 16 kpc in this example. The $\Delta t$s indicate the time elapsed from the beginning of star formation, in other words the various galactic ages.}
\end{figure}

In Figs.~10 and 11 we plot log(N/O) vs. log(N/H) + 12. The models
shown in Fig.~10 are the same as the lower log(N/O) ``family'' of models of
Fig.~9 (``bursting models'') computed
with MM yields. 
In Fig.~11 we show a prediction for an outer region of the MW disk
(in this case 16 kpc): the chemical evolution of the outer regions of the Milky Way disk 
are computed by assuming an inside-out formation of the disk, thus
having a lower star formation rate at any time than the solar vicinity and resembling the evolution of magellanic irregular galaxies (the model shown here
was computed with the same parameters as model A of 
Chiappini et al. 2001; see also Calura et al. 2003). 
These two figures suggest: a) both ``bursting'' models
and outer disks can explain the measured abundance ratios in DLAs; b) in ``bursting
models'' the low values of N/O will last for many
Gyrs while in outer regions of spiral disks the increase of N is faster (in Fig.~11
we indicate the elapsed time since the start of the SF at the corresponding
N/O value). For the models in Fig.~10 the N enrichment is slower. By instance,
for the short dashed curve, which represents a model 
computed with an intermediate value for the star formation efficiency (0.2 Gyr$^{-1}$),
we have that after roughly 35 Myrs the system reaches a log(N/O) of $-$2.7. 
After that, log(N/O) increases slowly achieving the first peak only after $\sim$3 Gyrs 
due to the low star formation efficiency and also to the fact that the first burst 
formed not many stars due to its short duration.  
This value stays almost constant until the second burst occurs then lowering 
log(N/O) again, in this case after 9 Gyrs from the start of the star formation. 
This means that for most of the time these systems will show a low log(N/O) and 
also an [O/Fe] ratio around 0.3-0.2dex in agreement with what is observed in DLAs\footnote
{This value is based essentially on O/Zn and Si/Zn abundance ratios. In fact, part of Fe 
in DLAs is likely to be in dust form, while O and Zn are not depleted in dust. For
Si only a mild depletion is expected (see discussions in Centurion et al. 2003a,b; 
Pettini et al.
2002 and Vladilo 2002).}. 
Our results suggest that DLAs could be explained
by systems which are similar to BCGs, with short duration bursts.
In this framework, even the ``low N/O group of DLAs'' can be 
explained as systems in which both massive
stars and LIMS are contributing to the ISM enrichment. 
It remains to be seen to what extent
the MM yields for N in LIMS will be increased if the HBB
would be taken into account.

In the scenario of Lanfranchi \& Matteucci (2003) or Calura et al. (2003), DLAs
with low log(N/O) are young systems, i.e., systems where the N coming from intermediate
mass stars did not have time yet to contribute to the ISM enrichment. 
This is a consequence of the fact that 
in both works the authors adopted vdHG yields. On the other hand, in the 
``bursting'' models presented 
here, computed with MM yields, the so called ``low log(N/O) DLAs'' are systems spanning 
a broad range of ages, as in this case the increase of N from LIMS is slow. 
However, not only we still have to 
explain DLAs systems with log(N/O) ratios similar to the ones observed in BCGs (log(N/O)$\sim-$1.6 dex), 
but we also have to explain the BCGs themselves.

This can be obtained with the same kind of models
but assuming that, in these systems, the initial burst of SF involved more mass or, 
in other words, lasted longer. The star formation efficiencies are of the
same order than in the models we used before to explain the ``low log(N/O) DLAs''
(larger values would produce produce too much O/H or N/H).
To illustrate this point we show in Fig.~12 a family of models similar
to the ones shown in Fig.~10, computed with MM, but in which the 
bursts of star formation lasted longer (0.8 Gyrs). 
In this case it is clear that the larger log(N/O) observed in BCGs can
be achieved. By varying parameters such as the burst duration and the number of bursts
we can thus explain the larger log(N/O) observed in BCGs and in some DLAs. 
With a longer burst, both the oxygen and the N/O ratio increase.
The relatively higher N is due to the fact that the star formation lasts longer
and thus the secondary N contribution is larger.
This figure shows that these models are also able to reproduce 
the observed C/O ratios in BCGs.

\begin{figure}
\centering
\includegraphics[width=8cm,angle=0]{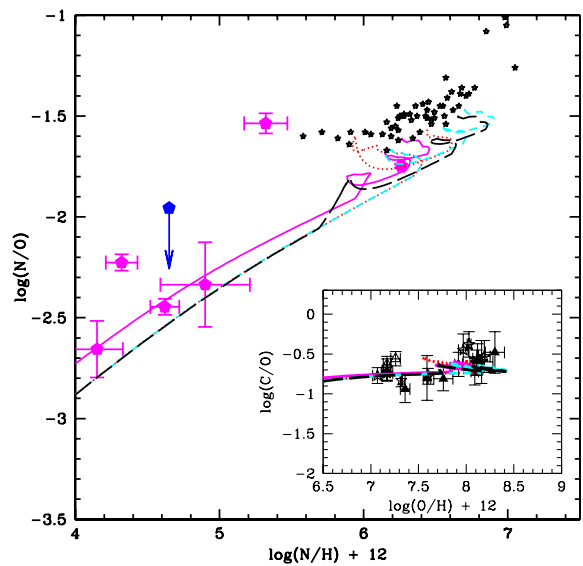}
\caption{Same as Fig.~10 but now we show model predictions for BCGs (see text). Also
shown is a log(C/O) vs. (O/H) plot. Details on the abundance data can be find in CRM2003.} 
\end{figure}

In summary, our results shows that once we adopt the MM yields, the low 
log(N/O) in DLAs can be explained
by ``bursting models'' of the kind of the ones suggested by
Lanfranchi \& Matteucci (2003). DLAs with low log(N/O) ratios would be 
systems similar to BCGs but in 
which the bursts of star formation were brief. 
In such models the N comes not only from massive stars but also from LIMS. 
Models with longer star formation bursts can
explain the abundance ratios of BCGs and DLAs with larger log(N/O) ratios. 
Notice that in this framework, the ``low'' and ``high'' log(N/O) DLAs would 
not have an age difference but rather a difference in their star formation mode.
Moreover, the difference between these two kind of systems will be better seen
in a log(N/O) vs. log(N/H) diagram as suggested by Centurion et al. (2003a). However,
at variance with what is suggested by these authors, 
we favor the view that the discontinuity seen in such a diagram indicates
a difference in the star formation rather than in the age. Our conclusions
depend on the adopted stellar yields and can change if either the amount of primary
N in massive stars is larger than the one adopted here or if HBB contributes to an
important amount of N in LIMS.

An anti-correlation between N/O and O/Fe would certainly help us to clarify
which is the best scenario for the formation of DLAs. In the scenario in which
DLAs with low log(N/O) are young systems, one would expect these objects to show
larger O/Fe ratios compared to systems with log(N/O) $\sim-$1.6 (as the BCGs).
However, this interpretation
is complicated by the fact that Fe is depleted onto dust grains.
In fact, accounting for the 
fraction of Fe in dust form requires taking into consideration the full complement
of abundances measurements in a DLA (see Vladilo 2002 and Pettini et al. 2002). 
Another discriminant would be s-process elements produced uniquely 
by intermediate mass stars (heavy s-elements).

\section{The Helium Enrichment}

Finally, in this last section we compare our results for the evolution
of helium in the MW predicted by models A and B. As shown in Sect.~2
the MM yields for helium are larger than the ones of vdHG and WW. 
In this work we decided to adopt the primordial composition from Chiappini et al. (2001), 
where for the pregalactic helium abundance we had $Y_p=$0.241.
In Fig.~13 we show the predictions of models A (solid line) and B (dashed line) 
for the evolution of the helium abundance as a function of metallicity (oxygen in this case). 
As it can be seen, model B is in better agreement with the solar value.
This model predicts a solar value of  $Y_{\odot} = $0.265 and a value of 0.272 
at the present time. Model A instead predicts
$Y_{\odot} =$0.253 and $Y=$0.257 at the present time.
As already
shown in Chiappini et al. (2002), when the yields of vdHG + WW are adopted the models
are marginaly consistent with the solar value.
Moreover, while models adopting vdHG+WW yields predict a $\Delta Y/\Delta Z$
value around 1.5, model B (with MM yields) predicts $\Delta Y/\Delta Z \sim$2.4.
This larger value for $\Delta Y/\Delta Z$ seems to be in better agreeement 
with the observations (see Pagel 2000).

\begin{figure}
\centering
\includegraphics[width=6cm,angle=0]{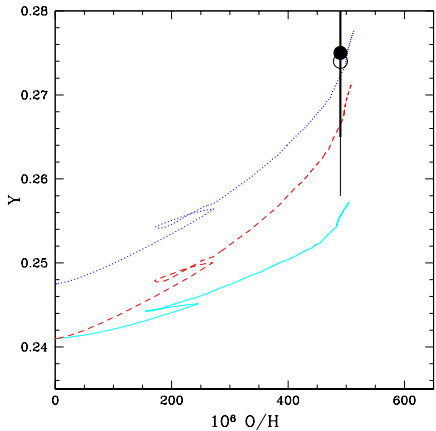}
\caption{$Y$ vs. 10$^6$O/H predicted by model A (solid line) and B 
(dashed line). The dotted-line show a model which is the same as model B
but computed with Yp =~0.248 (see text). The solar value
is also shown (where oxygen is from Allende-Prieto et al. 2001 and Y is from
Anders \& Grevesse 1989 - open circle with thin errorbar and Grevesse \& Sauval 1998 - black
circle with thick errorbar).}
\end{figure}

However, as shown in Chiappini et al. 2002, by means of our chemical evolution
model for the MW, we can constrain the primordial value of deuterium abundance and 
we find a value of (D/H)$_p \leq$ 4$\times$ 10$^{-5}$, which implies $Y_p >$ 0.244.
This is in fact in agreement with the recent results from WMAP which implies $Y_p =$0.248
(see Romano et al. 2003). In Fig.~13 we also show a model computed with the same nucleosynthesis
prescriptions of model B, but assuming $Y_p =$0.248 (dotted line). In this case
we obtain $Y_{\odot} = $0.272 in agreement with Grevesse \& Sauval (1998) and 
Bahcall et al. (2001). Model A computed with 
$Y_p =$0.248 gives $Y_{\odot} = $0.261 and it is still only marginally consistent
with the solar value.

\section{Conclusions}

\begin{itemize}

\item
We suggest, on the basis of the available data in the solar vicinity,
that C should come mainly from low- and intermediate-mass stars.
This is at variance with the interpretation by several authors (e.g.
Carigi 2000, Henry et al. 2000) that C should originate mainly in massive
stars. This conclusion was based on the yields of Maeder (1992) predicting
a strong metallicity dependence of the C produced in massive stars. However,
this work is now superseded by the new models of MM which
take into account stellar rotation effects and use weaker mass loss rates
accounting for the effects of clumping.
We show that with the new stellar yields of MM, C produced in massive stars is
not enough to explain the solar C/O ratio. Once the 3rd dredge up will be 
included in MM calculations, a rise in the C/O ratio will be obtained 
due to the contribution of LIMS. This is in fact seen if one adopts the
stellar yields of vdHG which include the 3rd dredge-up during the TP-AGB 
stellar phase. Moreover, the C yields from massive stars are underestimated 
in WW calculations (which do not account for mass loss by stellar winds
and do not include rotation) as already noticed
by Henry et al. (2000) and CRM2003. 

\item
Given our previous conclusion, we would expect that the ``gap'' already 
observed in the [Fe/O] vs. [O/H] and 
[Fe/Mg] vs. [Mg/H] diagrams (Gratton et al. 2000; Fuhrmann 1998) is 
also seen in the log(C/O) vs. log(O/H) plot. The homogeneous 
sample of Nissen (2003) seems to show the predicted discontinuity but
more data are needed in order to confirm this prediction.
The existence of such a ``gap'' is due to the halt in the star formation
between the end of the thick disk and beginning of the thin-disk phase and
in our model is naturally produced by the threshold in the star formation.

\item
The new yields of MM predict primary N production in massive stars. We show
that models for the MW computed with this new set of yields show
a plateau in log(N/O), due to massive stars with initial rotational velocities of 300 km$sec^{-1}$, at log(N/O) $\sim-4$. 
This value is below the value of $-$2.2 dex observed in some DLAs and hence we suggest
that in these systems both, massive and intermediate mass stars, are 
responsible for the N enrichment.
This is at variance with recent claims that massive stars were the only
ones to enrich systems which show a log(N/O)$\sim-$2.2 (Centurion et al. 2003a).

\item
When the MM yields are applied to the whole range
of masses, a slower increase of N with respect to what is obtained with
vdHG yields is found and this has important implications for the
interpretation of the DLAs abundance data. 
We suggest that DLAs are best reproduced by ``bursting models''.
Moreover, in this case the ``low'' and ``high'' log(N/O) DLAs
(if they really exist as separated groups) could be
explained as systems which show differences in their star formation history
rather than an age difference. In such a framework,
we are able to obtain systems which show both a low log(N/O) 
and a [O/Fe]$\sim$0.2-0.3 dex during almost all of their evolution. 
Outer regions of spirals can still explain the DLAs data (e.g. Calura
et al. 2003) but predict that
DLAs with low log(N/O) are quite young systems (younger than $\sim$150 Myr)
and this may happen as an odd coincidence.
However, we call attention to the fact that the primary N given 
by MM for the intermediate mass range should be seen as a lower 
limit as HBB can contribute to further increase this element.

\item
The new yields of MM for helium lead to a better agreement between the solar 
abundance value predicted by the MW models and the observed one. Moreover, 
when the MM yields are adopted we find $\Delta Y/\Delta Z \sim$ 2.4, whereas a value of 
1.5 is found by using vdHG+WW stellar yields. This shows that
although the $\Delta Y/\Delta Z$ values obtained in chemical evolution models
are almost independent on the primordial $Y_p$ adopted (see Chiappini et al. 2002), 
they do depend strongly on the adopted stellar yields.

\end{itemize}

\begin{acknowledgements}
We thank D. Romano, S. Recchi and M. Centurion for many useful
discussions.
We also thank the referee, R.B.C. Henry, for his suggestions
that improved this work.
\end{acknowledgements}

\end{document}